\documentclass{ieeeaccess}

\let\headerlogo\relax
\let\headerlogoall\relax

\makeatletter
\def\ps@headings{%
  \def\@oddhead{\parbox[t]{\textwidth}{\mbox{}\\[-6.3mm]{\raisebox{-2pt}{\headerfont\rightmark}}\hfill\headerlogoall\mbox{}\vspace*{-1.5mm}\par\hrulefill}}
  \def\@evenhead{\parbox[t]{\textwidth}{\mbox{}\\[-6.3mm]\mbox{}\headerlogoall\hfill{\raisebox{-2pt}{\headerfont\leftmark}}\vspace*{-1.5mm}\par\hrulefill}}
  \def\@oddfoot{\raisebox{8pt}{\hbox to 0pc{\hbox to \textwidth{\mbox{}\hfill{\footerpagefont\thepage}}}}}
  \def\@evenfoot{\raisebox{8pt}{\hbox to 0pc{\hbox to \textwidth{{\footerpagefont\thepage}\mbox{}\hfill}}}}
}
\def\ps@titlepage{%
  \def\@oddhead{\parbox[t]{\textwidth}{\mbox{}\\[-6.3mm]\mbox{}\nheaderfont\color{accessblue}\headName\hfill\headerlogo{\raisebox{-2pt}{\,}}\color{black}\vspace*{-1.5mm}\par\hrulefill}}
  \def\@evenhead{\parbox[t]{\textwidth}{\mbox{}\\[-6.3mm]\mbox{}\headerlogo\hfill\nheaderfont\color{accessblue}\headName{\raisebox{-2pt}{\,}}\color{black}\vspace*{-1.5mm}\par\hrulefill}}
  \def\@oddfoot{\raisebox{3pt}{\hbox to 0pc{\hbox to \textwidth{\mbox{}\hfill{\footerpagefont\thepage}}}}}
  \def\@evenfoot{\raisebox{3pt}{\hbox to 0pc{\hbox to \textwidth{{\footerpagefont\thepage}\hfill}\mbox{}}}}
}
\makeatother

\makeatletter
\long\def\@makecaption#1#2{%
\ifx\@captype\@IEEEtablestring%
  \begin{flushleft}
  \vspace*{5pt}
  {\vss\color{accessblue}\tablecapheadfont #1. \ }{\raggedright\tablecapfont#2\vss}%
  \end{flushleft}
  \@IEEEtablecaptionsepspace%
\else
  \@IEEEfigurecaptionsepspace%
  \setbox\@tempboxa\hbox{{\color{accessblue}\figcapheadfont #1. \ }}%
  {\vss\raggedright\noindent\unhbox\@tempboxa\figcapfont#2\vss}%
\fi\vskip 1pt plus 1pt minus 1pt}
\makeatother

\usepackage{cite}
\usepackage{amsmath,amssymb,amsfonts}
\usepackage{algorithmic}
\usepackage{booktabs}
\usepackage{graphicx}
\usepackage{textcomp}
\def\BibTeX{{\rm B\kern-.05em{\sc i\kern-.025em b}\kern-.08em
    T\kern-.1667em\lower.7ex\hbox{E}\kern-.125emX}}
\usepackage[hidelinks,bookmarks=false]{hyperref}

\begin{document}
\history{Preprint submitted to arXiv.}
\doi{}

\title{Resource-Efficient Hybrid Quantum Neighborhood Selection for Large-Scale Molecular Diversity Optimization}

\author{\uppercase{Nicolas Mendes de Ara\'{u}jo\authorrefmark{2,3}}
	\uppercase{and Lester de Abreu Faria\authorrefmark{1,2,3}}
}
\address[1]{Technological Institute of Aeronautics (ITA), Electronic Division, Mal Eduardo Gomes, 50, S\~{a}o Paulo -- SP, Brazil (e-mail: lester@ita.br)}
\address[2]{ICTQ Foton --- Institute of Quantum Sciences and Technologies Foton, Brazil}
\address[3]{LACQ Feynman --- Academic League of Quantum Computing (Liga Acad\^{e}mica de Computa\c{c}\~{a}o Qu\^{a}ntica Feynman), Brazil}

\markboth
{Preprint submitted to arXiv}
{Mendes de Ara\'{u}jo and Faria: Resource-Efficient Hybrid Quantum Neighborhood Selection}

\corresp{Corresponding author: Lester de Abreu Faria (e-mail: lester@ita.br).}

\begin{abstract}
	Large-scale combinatorial optimization remains computationally demanding even for mature classical heuristics, particularly when dense Quadratic Unconstrained Binary Optimization (QUBO) formulations induce large memory footprints, high CPU utilization, and long wall-clock execution times. While near-term quantum processors are not yet capable of delivering unconditional quantum advantage for such problems, hybrid quantum--classical architectures may provide practical value by reducing the computational resource burden associated with large-scale optimization workflows. This paper presents a resource-efficiency study of Hybrid Quantum Neighborhood Selection (HQNS), a hybrid quantum--classical framework that decomposes large dense QUBO instances into bounded-width quantum subproblems through stochastic frontier selection. The method is evaluated on the Maximum Diversity Subset Selection Problem (MDSSP), a representative NP-hard problem with relevance to molecular diversity selection. Rather than claiming strict dominance in final solution quality, the study focuses on the trade-off between solution quality retention and computational resource consumption. Experimental results on molecular diversity benchmarks up to $N=1000$ candidates show that HQNS preserves $99.9908\%$ of the mean diversity score achieved by an 11-restart parallel Simulated Annealing baseline while reducing wall-clock time by $94.91\%$, peak CPU utilization by $64.68\%$, and peak memory usage by $88.61\%$ relative to that strongest quality-oriented baseline. The QPU execution time remains bounded within an approximately $6$--$7$ second envelope across increasing problem scales, indicating that the quantum execution component is effectively decoupled from the global QUBO dimension when the frontier size is fixed. These results suggest that HQNS provides a resource-aware pathway for deploying hybrid quantum optimization in practical large-scale settings, not as evidence of unconditional quantum advantage, but as a computationally efficient architecture for incorporating near-term quantum processors into classical optimization pipelines.
\end{abstract}

\begin{keywords}
	Hybrid quantum--classical optimization, resource-efficient optimization, quantum optimization, QUBO, NISQ computing, molecular diversity selection, maximum diversity subset selection, simulated annealing, quantum neighborhood selection, stochastic frontier decomposition, computational resource efficiency, variational quantum algorithms.
\end{keywords}

\titlepgskip=-15pt

\maketitle

\section{Introduction}
\label{sec:introduction}

Large-scale combinatorial optimization remains a central computational bottleneck in science, engineering, logistics, finance, and molecular discovery. Many practically relevant problems can be formulated as Quadratic Unconstrained Binary Optimization (QUBO) instances, where the objective function depends on pairwise interactions among binary decision variables \cite{tabi_hybrid_2024,gurobi_2023,glover_qubo_survey_2022}. Although QUBO formulations provide a unified mathematical representation for a broad class of NP-hard optimization problems, dense interaction matrices impose substantial computational pressure on classical solvers, particularly in terms of memory footprint, CPU utilization, and wall-clock execution time.

The Maximum Diversity Subset Selection Problem (MDSSP) is a representative example of this challenge. Given a library of $N$ candidate elements and a pairwise similarity or dissimilarity matrix, the objective is to select a subset of size $K$ that maximizes aggregate diversity. In molecular discovery, this problem arises naturally when selecting chemically diverse compounds from large candidate libraries for downstream screening, commonly using molecular fingerprints and Tanimoto-type similarity measures \cite{rogers_ecfp_2010,valli_brnpdb_2023}. Since molecular similarity matrices are typically dense, the resulting QUBO representation scales quadratically with the number of candidates, creating a challenging benchmark for both classical and quantum optimization methods.

Classical heuristics such as Greedy search, Simulated Annealing (SA), multi-restart variants of SA, and commercial optimization solvers remain highly effective for MDSSP and related QUBO problems \cite{gurobi_2023,tang_dequantization_2019}. In particular, parallel multi-restart SA can achieve strong solution quality by exploring multiple regions of the search space simultaneously. However, this performance is obtained at the cost of increased CPU utilization, memory consumption, and total computational effort. In large-scale settings, the relevant question is therefore not only which method achieves the highest objective value, but also how much computational resource is required to obtain a solution of comparable quality.

Near-term quantum computing introduces a different optimization paradigm. Variational quantum algorithms, including the Variational Quantum Eigensolver (VQE) and the Quantum Approximate Optimization Algorithm (QAOA), provide a hybrid quantum--classical approach in which quantum circuits sample candidate states and classical optimizers update variational parameters \cite{peruzzo_vqe_2014,farhi_qaoa_2014}. However, direct QAOA encoding of dense QUBO instances remains impractical on noisy intermediate-scale quantum (NISQ) hardware because circuit width, depth, connectivity, and noise accumulation jointly constrain executable problem sizes \cite{preskill_nisq_2018,pagano_qaoa_2020,liu_layervqe_2022}. A monolithic formulation with $N$ binary variables requires quantum circuits whose width, connectivity, and two-qubit interaction structure rapidly exceed the limitations of current devices. Consequently, near-term quantum optimization is unlikely to deliver unconditional quantum advantage for dense large-scale QUBO problems in its direct form.

This limitation motivates a more practical research question: can hybrid quantum--classical methods provide computational utility by reducing resource consumption while preserving near-baseline solution quality? In this paper, we investigate this question through the lens of Hybrid Quantum Neighborhood Selection (HQNS), a resource-aware hybrid optimization framework that decomposes a large dense QUBO into a sequence of bounded-width quantum subproblems. Instead of encoding all $N$ variables into a single monolithic quantum circuit, HQNS selects a compact frontier of $F \ll N$ active variables, keeps the remaining variables classically frozen, and optimizes local neighborhoods through shallow quantum circuits. A stochastic frontier rotation mechanism then progressively explores different regions of the global solution landscape.

In this context, HQNS is not proposed as a monolithic quantum optimizer. It is a decomposition-based hybrid workflow in which the QPU is used as a bounded-width local optimizer over selected neighborhoods, while the classical controller maintains the global solution, updates the active frontier, and evaluates resource-level performance.

The central hypothesis of this work is that HQNS can preserve solution quality close to strong classical baselines while reducing the computational resource burden of large-scale optimization. This paper therefore does not frame HQNS as a strict quantum-advantage demonstration. Rather, it evaluates HQNS as a resource-efficient hybrid quantum--classical architecture. The focus is on measurable computational indicators: diversity-score retention, wall-clock time, peak CPU utilization, memory usage, GPU involvement, and QPU execution-time stability.

Using molecular diversity selection benchmarks derived from dense similarity matrices, we compare HQNS against classical Greedy search, single-restart Simulated Annealing, and 11-restart parallel Simulated Annealing. At the largest benchmark scale, $N=1000$, HQNS preserves $99.9908\%$ of the mean diversity score achieved by the 11-restart parallel SA baseline, while reducing wall-clock time by $94.91\%$, peak CPU utilization by $64.68\%$, and peak memory usage by $88.61\%$. In addition, the quantum execution component remains bounded within an approximately $6$--$7$ second envelope across increasing problem sizes, supporting the claim that the QPU workload is effectively decoupled from the global QUBO dimension when the frontier size is fixed.

The main contributions of this paper are summarized as follows:

\begin{itemize}
	\item We present HQNS as a resource-efficient hybrid quantum--classical optimization framework for dense large-scale QUBO problems.

	\item We evaluate HQNS on the Maximum Diversity Subset Selection Problem, a relevant NP-hard benchmark for molecular diversity selection.

	\item We quantify the quality--resource trade-off between HQNS and classical simulated annealing baselines, including wall-clock time, CPU utilization, memory consumption, and QPU execution time.

	\item We show that HQNS preserves $99.9908\%$ of the mean diversity score of an 11-restart parallel SA baseline at $N=1000$ while substantially reducing computational resource usage.

	\item We discuss the role of bounded-width quantum subproblems as a practical mechanism for incorporating NISQ processors into large-scale optimization pipelines without claiming unconditional quantum advantage.
\end{itemize}

The remainder of this paper is organized as follows. Section~II reviews related work on classical heuristics, quantum optimization, and resource-aware hybrid computing. Section~III formulates the molecular diversity selection problem and its QUBO representation. Section~IV describes the HQNS framework and its stochastic frontier decomposition strategy. Section~V presents the experimental methodology and resource-efficiency metrics. Section~VI reports the experimental results. Section~VII discusses implications, limitations, and deployment considerations. Finally, Section~VIII concludes the paper.

\section{Related Work}

\subsection{Classical Heuristics for Dense Combinatorial Optimization}

Dense combinatorial optimization problems remain challenging even when approximate methods are used. In the Maximum Diversity Subset Selection Problem (MDSSP), the objective depends on pairwise relationships among candidate elements, producing a dense similarity or dissimilarity matrix. As the number of candidates $N$ increases, both memory usage and computational effort grow rapidly, particularly when the algorithm repeatedly evaluates local changes in the objective function.

Classical heuristic methods such as Greedy search, Simulated Annealing (SA), Genetic Algorithms, Tabu Search, multi-start local search, and high-performance mathematical programming solvers have been widely used for such problems \cite{gurobi_2023,tang_dequantization_2019}. Greedy strategies are computationally efficient and often provide strong initial solutions, but they are susceptible to local traps caused by early irreversible selections. Simulated Annealing partially mitigates this issue by allowing stochastic transitions during the search process, while parallel or multi-restart SA improves exploration by executing multiple independent chains. However, these improvements typically increase CPU utilization, memory footprint, and total computational cost.

In practical large-scale settings, the strongest classical baseline is often not a single heuristic run, but a parallel or multi-restart version of that heuristic. This is important for fair evaluation: a hybrid quantum--classical method should not be compared only against weak classical baselines. In this work, we therefore compare HQNS not only against Greedy and single-restart SA, but also against an 11-restart parallel SA baseline, which provides a more demanding reference point for solution quality and resource consumption.

\subsection{Quantum Optimization and NISQ Constraints}

Quantum optimization methods have attracted substantial interest due to their potential to sample complex combinatorial landscapes using quantum states. Among these methods, QAOA is one of the most prominent variational approaches for solving QUBO and Ising-type problems \cite{farhi_qaoa_2014,zhou_hybrid_access_2020}, building on the broader hybrid variational paradigm introduced by early VQE demonstrations \cite{peruzzo_vqe_2014} and reviewed comprehensively in \cite{cerezo_vqa_review_2021}. In principle, QAOA maps a binary optimization problem into a parameterized quantum circuit whose measurement outcomes correspond to candidate solutions.

However, direct QAOA application to dense QUBO instances faces severe limitations on noisy intermediate-scale quantum (NISQ) processors. A dense QUBO with $N$ variables induces $O(N^2)$ pairwise interactions, which becomes especially problematic under NISQ constraints and sparse hardware connectivity \cite{preskill_nisq_2018,pagano_qaoa_2020,harrigan_quantum_access_2021}. Implementing these interactions requires a large number of two-qubit gates, and mapping them onto sparse hardware connectivity introduces additional routing and transpilation overhead. As a result, circuit depth and error accumulation become limiting factors long before problem sizes of practical industrial relevance are reached.

These limitations have motivated alternative hybrid strategies that reduce quantum circuit width and depth while retaining some capacity for structured exploration, including layer-wise variational methods, coordinate-descent-inspired quantum optimization, warm-starting, and nonnative hybrid quantum--classical formulations \cite{liu_layervqe_2022,dupont_quacod_2023,egger_warmstart_2021,tabi_hybrid_2024}. Such strategies include problem decomposition, warm-starting, circuit cutting, local neighborhood optimization, scaling analysis \cite{pelofske_scaling_access_2023}, and hybrid workflows in which classical computation performs global coordination while the quantum processor handles compact subproblems. HQNS follows this general direction by restricting each quantum execution to a small frontier of $F \ll N$ active variables, while the remaining variables are held fixed and incorporated into the reduced QUBO coefficients.

\subsection{Resource-Aware Hybrid Quantum--Classical Computing}

Most studies of quantum optimization emphasize final objective value, approximation ratio, or convergence behavior, although training cost, optimizer efficiency, and noise-aware robustness have become increasingly important in recent variational quantum algorithm studies \cite{shaydulin_symmetry_2021,tibaldi_bayesian_2023,he_shiftednoise_2024}. While these metrics remain important, they do not fully characterize practical computational value. For near-term hybrid quantum--classical systems, resource-efficiency metrics are equally relevant, including wall-clock time, CPU utilization, memory usage, GPU involvement, QPU execution time, and the stability of these quantities as problem size increases \cite{preskill_nisq_2018,shaydulin_symmetry_2021}.

This resource-aware perspective is particularly important because hybrid quantum algorithms do not replace classical computation; they reorganize it. The total workflow typically includes classical preprocessing, circuit construction, parameter optimization, quantum sampling, postprocessing, and solution reconstruction. Therefore, a meaningful evaluation must account for the complete hybrid pipeline rather than only the quantum circuit execution time.

In this context, the value of a NISQ-compatible algorithm may not appear as strict superiority in objective value. Instead, practical utility may emerge when a hybrid method preserves near-baseline solution quality while reducing computational resource consumption or enabling workflows that would be impractical under monolithic quantum encodings. This paper adopts this viewpoint and evaluates HQNS as a resource-efficient hybrid optimization framework. The central question is not whether HQNS achieves unconditional quantum advantage, but whether it provides a favorable quality--resource trade-off relative to strong classical baselines.

\subsection{Positioning of This Work}

This work differs from conventional quantum optimization benchmarking in three ways. First, it evaluates HQNS against a strong parallel classical baseline rather than only against Greedy or single-run heuristics. Second, it reports not only solution quality, but also wall-clock time, CPU utilization, memory footprint, and QPU execution-time stability. Third, it frames the contribution as resource-aware quantum utility under NISQ constraints, avoiding claims of unconditional quantum advantage.

The resulting perspective is aligned with practical deployment: if a hybrid quantum--classical method can preserve nearly all of the solution quality of a strong classical baseline while significantly reducing computational resource requirements and maintaining bounded quantum circuit width, it may provide operational value even without demonstrating asymptotic quantum superiority.

\section{Problem Formulation}

\subsection{Maximum Diversity Subset Selection Problem}

The Maximum Diversity Subset Selection Problem (MDSSP) consists of selecting a subset of $K$ elements from a larger candidate library of size $N$ such that the selected subset maximizes aggregate pairwise diversity. Let

\begin{equation}
	\mathcal{M} = \{m_1, m_2, \ldots, m_N\}
\end{equation}
denote a candidate library, where each element $m_i$ represents a molecular compound. Each pair $(m_i,m_j)$ is associated with a similarity coefficient $S_{ij} \in [0,1]$, where larger values indicate greater molecular similarity. In this work, $S_{ij}$ is computed using the Tanimoto coefficient over molecular fingerprints, following standard cheminformatics practice for extended-connectivity molecular representations \cite{rogers_ecfp_2010}:

\begin{equation}
	S_{ij} =
	\frac{|FP_i \cap FP_j|}
	{|FP_i \cup FP_j|}.
\end{equation}

The selection decision is represented by a binary vector

\begin{equation}
	x \in \{0,1\}^{N},
\end{equation}
where $x_i = 1$ indicates that molecule $m_i$ is selected and $x_i = 0$ otherwise. The cardinality constraint requires exactly $K$ selected molecules:

\begin{equation}
	\sum_{i=1}^{N} x_i = K.
\end{equation}

Since the objective is to maximize molecular diversity, the optimization problem can equivalently be written as the minimization of aggregate pairwise similarity among selected molecules:

\begin{equation}
	\min_{x}
	\sum_{i<j} S_{ij} x_i x_j
	\quad
	\mathrm{s.t.}
	\quad
	\sum_{i=1}^{N} x_i = K.
	\label{eq:mdssp}
\end{equation}

The corresponding diversity score is defined as

\begin{equation}
	D(x) =
	\frac{K(K-1)}{2}
	-
	\sum_{i<j} S_{ij}x_i x_j.
	\label{eq:diversity_score}
\end{equation}

A higher value of $D(x)$ indicates a more diverse selected subset. The theoretical maximum,

\begin{equation}
	D_{\max} = \frac{K(K-1)}{2},
\end{equation}
would occur only if all selected pairs were completely dissimilar, i.e., $S_{ij}=0$ for every selected pair. In real molecular libraries, this condition is rarely achieved due to shared scaffolds, conserved substructures, and chemical similarity among compounds. As a result, MDSSP provides a realistic dense combinatorial benchmark for evaluating both solution quality and computational resource efficiency.

\subsection{QUBO Representation}

To make the problem compatible with both classical and quantum optimization workflows, the constrained formulation in Eq.~\eqref{eq:mdssp} is transformed into a Quadratic Unconstrained Binary Optimization (QUBO) problem. The cardinality constraint is incorporated through a quadratic penalty term:

\begin{equation}
	H_{\mathrm{cost}}(x)
	=
	\sum_{i<j} S_{ij}x_i x_j
	+
	\lambda
	\left(
	\sum_{i=1}^{N}x_i - K
	\right)^2,
	\label{eq:qubo}
\end{equation}
where $\lambda > 0$ is a penalty coefficient controlling the enforcement of the subset-size constraint. Expanding Eq.~\eqref{eq:qubo} yields a standard QUBO form:

\begin{equation}
	H_{\mathrm{cost}}(x)
	=
	\sum_i Q_{ii}x_i
	+
	\sum_{i<j} Q_{ij}x_i x_j
	+
	C,
	\label{eq:qubo_expanded}
\end{equation}
where $Q_{ii}$ and $Q_{ij}$ are the linear and quadratic QUBO coefficients, respectively, and $C$ is a constant offset. The dense nature of the similarity matrix $S$ induces a dense QUBO interaction graph, since most molecular pairs contribute nonzero pairwise similarity terms.

This dense structure is central to the resource-efficiency analysis in this paper. For classical solvers, dense QUBO matrices increase memory usage and the cost of repeated objective evaluations. For monolithic quantum formulations, the same dense structure requires a large number of two-qubit interaction terms, creating circuits whose width, depth, and connectivity requirements rapidly exceed the limits of current NISQ hardware.

\subsection{Ising Mapping for Quantum Optimization}

For quantum optimization, the binary variables can be mapped to Pauli-$Z$ operators using the standard transformation

\begin{equation}
	x_i = \frac{1-Z_i}{2}.
\end{equation}

Under this substitution, the QUBO Hamiltonian becomes an Ising-type Hamiltonian:

\begin{equation}
	H =
	\sum_i h_i Z_i
	+
	\sum_{i<j} J_{ij} Z_i Z_j
	+
	C',
	\label{eq:ising}
\end{equation}
where $h_i$ are local field coefficients, $J_{ij}$ are pairwise coupling coefficients, and $C'$ is a constant energy shift.

A direct QAOA implementation of Eq.~\eqref{eq:ising} would require a quantum register with $N$ qubits and a circuit layer containing $O(N^2)$ two-qubit interaction terms for dense QUBO instances. This makes monolithic QAOA impractical for the larger benchmark cases considered in this paper, such as $N=500$ and $N=1000$. The HQNS framework addresses this issue by avoiding full-system quantum encoding. Instead, it constructs reduced QUBO subproblems over compact active frontiers of size $F \ll N$, allowing the quantum circuit width to remain bounded even as the global problem size increases.

\subsection{Quality and Resource-Efficiency Objectives}

The evaluation in this paper considers two complementary objectives. The first is solution quality, measured by the diversity score $D(x)$ in Eq.~\eqref{eq:diversity_score}. The second is computational resource efficiency, measured through wall-clock execution time, peak CPU utilization, peak memory usage, GPU involvement, and QPU execution time.

To compare HQNS against a strong classical baseline, we define the diversity retention ratio:

\begin{equation}
	R_D =
	\frac{\overline{D}_{\mathrm{HQNS}}}
	{\overline{D}_{\mathrm{SA}}}
	\times 100\%,
	\label{eq:diversity_retention}
\end{equation}
where $\overline{D}_{\mathrm{HQNS}}$ is the mean diversity score obtained by HQNS over multiple independent runs and $\overline{D}_{\mathrm{SA}}$ is the corresponding mean diversity score obtained by the 11-restart parallel Simulated Annealing baseline.

Similarly, the relative reduction in a computational resource $r$ is defined as

\begin{equation}
	\Delta_r =
	\left(
	1 -
	\frac{\overline{r}_{\mathrm{HQNS}}}
	{\overline{r}_{\mathrm{SA}}}
	\right)
	\times 100\%,
	\label{eq:resource_reduction}
\end{equation}
where $r$ may represent wall-clock time, peak CPU utilization, or peak memory usage. These metrics allow the comparison to focus not only on final solution quality, but also on the computational cost required to obtain that quality.

This formulation supports the central evaluation criterion of the paper: HQNS is considered practically useful if it preserves near-baseline solution quality while substantially reducing the resource footprint of the optimization workflow.

\section{Hybrid Quantum Neighborhood Selection Framework}

\subsection{Overview}

Hybrid Quantum Neighborhood Selection (HQNS) is a hybrid quantum--classical optimization framework designed to make large dense QUBO instances tractable under near-term quantum hardware constraints. Rather than encoding the full $N$-variable problem into a monolithic quantum circuit, HQNS decomposes the global optimization task into a sequence of compact neighborhood subproblems. Each subproblem is defined over an active frontier of $F \ll N$ decision variables, while the remaining variables are kept fixed and treated classically.

This design has two practical consequences. First, the quantum circuit width depends on the frontier size $F$, rather than on the full problem size $N$. Second, the dense global QUBO is never fully executed as a quantum circuit. Instead, the quantum processor is used as a bounded-width local optimizer embedded within a classical coordination loop. This makes HQNS suitable for resource-aware hybrid optimization, where the goal is not only to obtain high-quality solutions, but also to reduce wall-clock time, CPU usage, memory footprint, and quantum circuit complexity.

The general HQNS workflow consists of five main steps:

\begin{enumerate}
	\item Generate an initial feasible solution using a classical warm-start heuristic.
	\item Select an active frontier of $F$ variables based on marginal contribution scores.
	\item Construct a reduced QUBO over the active frontier while absorbing the influence of frozen variables into shifted coefficients.
	\item Optimize the reduced QUBO using a shallow QAOA-inspired variational circuit.
	\item Update the global solution and repeat the process through stochastic frontier rotation.
\end{enumerate}

This process allows HQNS to progressively refine a global solution through a sequence of small quantum-assisted neighborhood optimizations.

\subsection{Warm-Start Initialization}

HQNS begins with a feasible classical solution $x^{(0)} \in \{0,1\}^{N}$ satisfying the cardinality constraint

\begin{equation}
	\sum_{i=1}^{N} x_i^{(0)} = K.
\end{equation}

In this work, the initial solution is obtained using a deterministic greedy heuristic that iteratively selects molecules with high marginal diversity contribution. The role of the greedy initialization is not to solve the problem optimally, but to provide a valid and computationally inexpensive starting point for the hybrid refinement procedure.

The warm-start solution also reduces the burden placed on the quantum circuit. Instead of requiring the quantum processor to explore the full combinatorial search space from an unbiased superposition, HQNS focuses quantum sampling on local neighborhoods around an already feasible candidate solution. This is important from a resource-efficiency perspective because it reduces the number of quantum variables and optimization iterations required to obtain competitive solutions.

\subsection{Stochastic Frontier Selection}

At each optimization stage $s$, HQNS selects an active frontier $A_s \subset \{1,\ldots,N\}$ containing $F$ variables. These variables define the local neighborhood to be optimized by the quantum subroutine. The remaining variables form the frozen complement

\begin{equation}
	F_s = \{1,\ldots,N\} \setminus A_s.
\end{equation}

The frontier is selected using a marginal impact score that estimates the contribution of each variable to the current solution structure. Given the current solution $x^{(s)}$, the marginal impact of variable $i$ is defined as

\begin{equation}
	\mu_i(x^{(s)}) =
	\left|
	\sum_{j \neq i} S_{ij} x_j^{(s)}
	\right|.
	\label{eq:marginal_impact}
\end{equation}

Variables with high $\mu_i$ are strongly coupled to the current selected subset and are therefore relevant candidates for local modification. To avoid deterministic repetition of the same neighborhood, HQNS converts the marginal impact scores into a probability distribution:

\begin{equation}
	P(i \mid x^{(s)}) =
	\frac{
		\exp\left(\mu_i(x^{(s)})/\tau\right)
	}{
		\sum_{k=1}^{N}
		\exp\left(\mu_k(x^{(s)})/\tau\right)
	},
	\label{eq:frontier_distribution}
\end{equation}
where $\tau > 0$ is a temperature parameter controlling the exploration--exploitation trade-off. Lower values of $\tau$ concentrate probability mass on high-impact variables, while larger values promote broader exploration of the search space.

The active frontier $A_s$ is then sampled without replacement according to Eq.~\eqref{eq:frontier_distribution}, optionally balancing removal candidates from the currently selected subset and insertion candidates from the unselected pool. This stochastic rotation of frontiers enables HQNS to explore multiple local neighborhoods while preserving a bounded quantum circuit width.

\subsection{Reduced QUBO Construction}

Once the active frontier $A_s$ is selected, HQNS constructs a reduced QUBO involving only the $F$ active variables. The frozen variables are not discarded; instead, their influence is absorbed into the linear terms of the reduced QUBO.

Starting from the full QUBO

\begin{equation}
	H_{\mathrm{cost}}(x)
	=
	\sum_i Q_{ii}x_i
	+
	\sum_{i<j} Q_{ij}x_i x_j
	+
	C,
\end{equation}
the reduced QUBO at stage $s$ is written as

\begin{equation}
	H_s(x_{A_s})
	=
	\sum_{a \in A_s} \widetilde{Q}_{aa}x_a
	+
	\sum_{\substack{a,b \in A_s \\ a<b}}
	\widetilde{Q}_{ab}x_a x_b
	+
	\widetilde{C},
	\label{eq:reduced_qubo}
\end{equation}
where the shifted linear coefficient is

\begin{equation}
	\widetilde{Q}_{aa}
	=
	Q_{aa}
	+
	\sum_{\substack{f \in F_s \\ x_f^{(s)}=1}}
	Q_{af}.
	\label{eq:shifted_linear}
\end{equation}

The quadratic terms within the active frontier are preserved as

\begin{equation}
	\widetilde{Q}_{ab} = Q_{ab}, \quad a,b \in A_s.
\end{equation}

This reduction preserves the local interaction structure between the active variables and the frozen solution context. From a resource-efficiency viewpoint, Eq.~\eqref{eq:reduced_qubo} is the key transformation: the global dense QUBO remains classically represented, but only a compact $F$-variable subproblem is passed to the quantum circuit at each stage.

\subsection{Quantum Subproblem Optimization}

The reduced QUBO is mapped to an Ising Hamiltonian over $F$ qubits using the standard binary-to-spin transformation

\begin{equation}
	x_a = \frac{1-Z_a}{2}.
\end{equation}

The resulting Hamiltonian has the form

\begin{equation}
	H_s =
	\sum_{a \in A_s} h_a Z_a
	+
	\sum_{\substack{a,b \in A_s \\ a<b}}
	J_{ab} Z_a Z_b
	+
	C_s.
\end{equation}

HQNS then applies a shallow QAOA-inspired variational circuit to the reduced Hamiltonian \cite{farhi_qaoa_2014}:

\begin{equation}
	|\psi(\gamma,\beta)\rangle =
	\prod_{\ell=1}^{p}
	\left(
	e^{-i\beta_{\ell}H_M}
	e^{-i\gamma_{\ell}H_s}
	\right)
	|\psi_0\rangle,
	\label{eq:hqns_ansatz}
\end{equation}
where $H_M = \sum_{a \in A_s} X_a$ is the mixer Hamiltonian and $p$ is the circuit depth parameter. In this work, we use shallow circuits with $p=1$, consistent with the constraints of NISQ hardware.

The initial quantum state $|\psi_0\rangle$ is biased by the classical warm-start solution:

\begin{equation}
	|\psi_0\rangle =
	\bigotimes_{a \in A_s}
	R_y(\theta_a)|0\rangle,
\end{equation}
with

\begin{equation}
	\theta_a =
	\begin{cases}
		0.85\pi, & \text{if } a \in x^{(0)}, \\
		0.15\pi, & \text{otherwise}.
	\end{cases}
\end{equation}

This initialization concentrates probability amplitude around the classical seed while preserving variational freedom for local improvement, following the general rationale of warm-started quantum optimization \cite{egger_warmstart_2021}.

\subsection{Classical Parameter Optimization and CVaR Filtering}

The circuit parameters are optimized using Simultaneous Perturbation Stochastic Approximation (SPSA), a gradient-free stochastic optimization method well suited to noisy and low-dimensional variational loops \cite{spall_spsa_1998}. SPSA is well suited for hybrid quantum optimization because it estimates gradient directions using only two objective evaluations per iteration, independently of the number of variational parameters. This is relevant because parameter training can dominate the cost of QAOA-like workflows \cite{shaydulin_symmetry_2021,tibaldi_bayesian_2023}.

To improve robustness against noisy samples and low-quality bitstrings, HQNS uses a Conditional Value-at-Risk (CVaR) objective, following prior work on CVaR-based variational quantum optimization \cite{barkoutsos_cvar_2020}. Given measured bitstrings and their associated QUBO costs $C(x)$, the CVaR objective is defined as

\begin{equation}
	L_{\alpha}(\theta)
	=
	\mathbb{E}
	\left[
		C(x)
		\mid
		C(x) \leq q_{\alpha}
		\right],
	\label{eq:cvar}
\end{equation}
where $q_{\alpha}$ is the $\alpha$-quantile of the sampled cost distribution. Smaller values of $\alpha$ focus the optimization on the best-performing sampled bitstrings, improving robustness at the cost of higher estimator variance.

\subsection{Global Solution Update}

After optimizing the reduced QUBO, the best sampled frontier assignment is inserted back into the global solution. Let $b_s^{\star}$ denote the best frontier bitstring obtained at stage $s$. The global solution is updated as

\begin{equation}
	x_i^{(s+1)} =
	\begin{cases}
		b_{s,i}^{\star}, & i \in A_s, \\
		x_i^{(s)},       & i \in F_s.
	\end{cases}
\end{equation}

The procedure is repeated for $S$ stages, with each stage selecting a new stochastic frontier. This creates a multi-stage local refinement process in which the quantum processor repeatedly optimizes compact subproblems while the classical controller maintains the global solution state.

\subsection{Operational Resource Profile}

The resource-efficiency of HQNS arises from the separation between global problem representation and local quantum execution. The full similarity matrix is maintained classically, but the quantum processor receives only reduced subproblems of size $F$. As a result, the quantum circuit width is bounded by $F$, and the number of two-qubit interaction terms per quantum stage scales as

\begin{equation}
	O(F^2),
\end{equation}
instead of the

\begin{equation}
	O(N^2)
\end{equation}
interaction burden required by monolithic QAOA on the full dense QUBO.

The total hybrid workflow still includes classical preprocessing, frontier selection, reduced QUBO construction, SPSA iterations, and solution updates. For $S$ stages and $M$ SPSA iterations per stage, the overall computational cost can be expressed as

\begin{equation}
	T_{\mathrm{HQNS}}
	=
	O\left(S(N + MF^2)\right).
\end{equation}

Thus, HQNS does not eliminate classical computation. Instead, it reorganizes the optimization workload so that the quantum processor handles compact local subproblems while the classical system performs coordination, memory management, and global state updates.

This architecture is particularly relevant for resource-aware optimization. If $F$ and $M$ remain bounded, the QPU execution burden is effectively decoupled from the global QUBO dimension, while the classical overhead scales primarily with frontier selection and data access. This provides the basis for the quality--resource trade-off evaluated in the following sections.

\section{Experimental Methodology}

\subsection{Experimental Objective}

The experimental methodology is designed to evaluate HQNS as a resource-efficient hybrid quantum--classical optimization framework. The objective is not to demonstrate unconditional quantum advantage or strict dominance in final solution quality. Instead, the experiments assess whether HQNS can preserve near-baseline diversity performance while reducing computational resource usage relative to strong classical heuristics.

The evaluation therefore considers two complementary dimensions:

\begin{itemize}
	\item \textbf{Solution quality:} measured by the final molecular diversity score and QUBO cost.
	\item \textbf{Computational resource efficiency:} measured by wall-clock time, peak CPU utilization, memory footprint, GPU utilization, and QPU execution time.
\end{itemize}

This methodology supports a quality--resource trade-off analysis, which is central to the practical deployment of hybrid quantum--classical optimization in near-term computing environments.

\subsection{Benchmark Dataset}

The experiments are conducted on molecular diversity selection instances derived from molecular fingerprint similarity matrices, with molecular candidates obtained from the Brazilian Natural Product Database (BrNPDB) \cite{valli_brnpdb_2023}. Each candidate molecule is represented by a molecular fingerprint, and pairwise similarities are computed using the Tanimoto coefficient. The resulting dense similarity matrix defines the MDSSP instance and its corresponding QUBO formulation.

Six benchmark scales are considered:

\begin{equation}
	N \in \{30, 60, 120, 250, 500, 1000\}.
\end{equation}

The subset size $K$ is selected according to the problem scale, producing increasingly large combinatorial search spaces. Table~\ref{tab:benchmark_datasets_access} summarizes the benchmark configurations.

\begin{table}[t]
	\centering
	\caption{Benchmark datasets and search-space complexity.}
	\label{tab:benchmark_datasets_access}
	\resizebox{\columnwidth}{!}{
		\begin{tabular}{lccc}
			\toprule
			Dataset & $N$  & $K$ & Search Space $\binom{N}{K}$ \\
			\midrule
			Small-A & 30   & 8   & $5.9 \times 10^{6}$         \\
			Small-B & 60   & 10  & $7.5 \times 10^{10}$        \\
			Medium  & 120  & 10  & $2.3 \times 10^{13}$        \\
			Large-A & 250  & 15  & $2.1 \times 10^{22}$        \\
			Large-B & 500  & 30  & $1.8 \times 10^{56}$        \\
			Mega    & 1000 & 50  & $2.6 \times 10^{85}$        \\
			\bottomrule
		\end{tabular}
	}
\end{table}

The $N=1000$ benchmark is used as the main large-scale evaluation case for statistical validation and resource-efficiency analysis.

\subsection{Compared Optimization Methods}

HQNS is compared against three classical baselines, following the principle that hybrid quantum methods should be evaluated against strong classical and quantum-inspired references rather than only weak heuristic baselines \cite{tang_dequantization_2019,gurobi_2023}:

\begin{itemize}
	\item \textbf{Greedy Search:} a deterministic marginal-diversity heuristic used as a fast baseline and warm-start reference.

	\item \textbf{Simulated Annealing (Single):} a single stochastic annealing chain initialized from the candidate solution space.

	\item \textbf{Simulated Annealing Parallel-11:} eleven independent Simulated Annealing chains executed in parallel, representing a strong CPU-parallel classical baseline.
\end{itemize}

The SA Parallel-11 baseline is the main reference method for large-scale comparison, because multi-restart classical heuristics provide a demanding quality-oriented reference for assessing whether a hybrid quantum workflow offers practical utility \cite{tang_dequantization_2019}. This choice avoids an artificially weak benchmark and allows HQNS to be evaluated against a high-quality classical heuristic with substantial computational resource usage.

The HQNS configuration used for the main $N=1000$ benchmark is:
\begin{equation}
	S = 4, \quad M = 30, \quad F = 20, \quad \alpha = 0.05, \quad p = 1,
\end{equation}
where $S$ is the number of frontier-crawling stages, $M$ is the number of SPSA iterations, $F$ is the active frontier size, $\alpha$ is the CVaR filtering parameter, and $p$ is the QAOA-inspired circuit depth.

\subsection{Statistical Validation Protocol}

To evaluate run-to-run stability, the $N=1000$ benchmark is repeated over ten independent executions for all evaluated methods: Greedy, SA Single, SA Parallel-11, and HQNS. Each execution uses a distinct random seed when applicable. For deterministic components, repeated executions are retained to preserve a uniform reporting structure across all methods.

For each run, the following quantities are recorded:

\begin{itemize}
	\item final diversity score;
	\item final QUBO cost;
	\item validity of the cardinality constraint;
	\item wall-clock time;
	\item peak CPU utilization;
	\item peak memory usage;
	\item GPU utilization, when applicable;
	\item QPU execution time, when applicable.
\end{itemize}

For each method, the mean and standard deviation are computed for solution quality and computational resource metrics. This protocol allows the comparison to distinguish between pure solution quality, runtime efficiency, CPU and memory footprint, and QPU execution behavior.






\subsection{Solution-Quality Metrics}

The primary solution-quality metric is the diversity score $D(x)$:

\begin{equation}
	D(x) =
	\frac{K(K-1)}{2}
	-
	\sum_{i<j} S_{ij}x_i x_j.
\end{equation}

Higher values of $D(x)$ indicate more diverse selected subsets. The QUBO cost is also reported as a complementary metric, where lower values indicate better minimization of pairwise similarity and penalty terms.

To compare HQNS with SA Parallel-11, the diversity retention ratio is defined as:

\begin{equation}
	R_D =
	\frac{\overline{D}_{\mathrm{HQNS}}}
	{\overline{D}_{\mathrm{SA}}}
	\times 100\%,
	\label{eq:diversity_retention_access}
\end{equation}
where $\overline{D}_{\mathrm{HQNS}}$ and $\overline{D}_{\mathrm{SA}}$ denote the mean diversity scores across independent runs.

A value of $R_D$ close to $100\%$ indicates that HQNS preserves nearly all of the solution quality achieved by the classical parallel SA baseline.

\subsection{Computational Resource Metrics}

The resource-efficiency evaluation considers four main computational indicators.

\subsubsection{Wall-Clock Time}

Wall-clock time measures the total elapsed time required to complete the optimization workflow. For HQNS, this includes classical preprocessing, frontier selection, reduced QUBO construction, variational parameter optimization, quantum execution, and solution update. For SA Parallel-11, this includes the full parallel annealing process and final solution selection.

\subsubsection{Peak CPU Utilization}

Peak CPU utilization measures the maximum observed CPU load during execution. Since SA Parallel-11 uses multiple independent annealing chains, it can impose substantially higher CPU demand. HQNS, by contrast, shifts part of the optimization workload into bounded quantum subproblems and reduced local neighborhoods.

\subsubsection{Peak Memory Usage}

Peak memory usage captures the maximum memory footprint observed during optimization. Dense similarity matrices and repeated local search evaluations can increase memory pressure in classical heuristics. Reducing memory footprint is practically relevant because memory is increasingly expensive in modern AI and high-performance computing environments.

\subsubsection{QPU Execution Time}

For HQNS, QPU execution time measures the time spent executing the quantum subproblem on the quantum processing unit. Since HQNS operates on fixed-size frontiers, QPU execution time is expected to remain approximately stable as the global problem size $N$ increases, provided that the frontier size $F$ remains bounded.

\subsection{Resource-Reduction Metrics}

For each computational resource $r$, the relative reduction achieved by HQNS with respect to SA Parallel-11 is computed as:
\begin{equation}
	\Delta_r =
	\left(
	1 -
	\frac{\overline{r}_{\mathrm{HQNS}}}
	{\overline{r}_{\mathrm{SA}}}
	\right)
	\times 100\%.
	\label{eq:resource_reduction_access}
\end{equation}

In this work, $r$ represents wall-clock time, peak CPU utilization, and peak memory usage. Positive values of $\Delta_r$ indicate that HQNS uses fewer computational resources than the SA Parallel-11 baseline.

The resulting quality--resource trade-off is evaluated by jointly considering $R_D$ and $\Delta_r$. A practically favorable outcome occurs when HQNS achieves high diversity retention while producing substantial reductions in computational resource usage.

\subsection{Hardware and Execution Environment}

All classical and hybrid executions are performed under the same computational environment to ensure comparability of wall-clock time, CPU usage, and memory measurements. The HQNS workflow uses GPU acceleration only for parallel numerical operations when available, while quantum execution is restricted to compact frontier subproblems.

The QPU component is evaluated through bounded-width quantum executions associated with the reduced HQNS frontier Hamiltonians. Since the frontier size remains fixed or slowly varying across problem scales, the QPU workload is expected to remain stable even as the global MDSSP instance increases from $N=250$ to $N=1000$.

The resource measurements are interpreted as empirical workflow-level indicators. They are not direct electrical energy measurements. Therefore, this paper reports CPU, memory, GPU, QPU time, and wall-clock metrics as computational resource proxies rather than direct power-consumption measurements.

\subsection{Evaluation Criteria}

The experimental evaluation is guided by the following criteria:

\begin{enumerate}
	\item \textbf{Quality retention:} HQNS should preserve a high percentage of the mean diversity score achieved by SA Parallel-11.

	\item \textbf{Time reduction:} HQNS should reduce wall-clock execution time relative to SA Parallel-11.

	\item \textbf{CPU reduction:} HQNS should reduce peak CPU utilization relative to SA Parallel-11.

	\item \textbf{Memory reduction:} HQNS should reduce peak memory usage relative to SA Parallel-11.

	\item \textbf{QPU stability:} HQNS should maintain bounded QPU execution time as $N$ increases, provided that the frontier size $F$ remains bounded.
\end{enumerate}

These criteria reflect the central premise of this paper: the practical value of HQNS lies not in strict objective-value dominance over classical heuristics, but in its ability to deliver near-baseline solution quality with a lower computational resource footprint and a NISQ-compatible execution structure.

\section{Experimental Results}

\subsection{Solution Quality Retention}

This section evaluates whether HQNS preserves solution quality relative to a strong classical baseline. The main comparison is performed at the largest benchmark scale, $N=1000$ and $K=50$, using ten independent runs for both HQNS and the 11-restart parallel Simulated Annealing baseline.

Table~\ref{tab:quality_all_methods} reports the statistical comparison of final diversity scores and QUBO costs. SA Parallel-11 achieves the higher mean diversity score, with $\overline{D}_{\mathrm{SA}} = 1160.9157$, while HQNS achieves $\overline{D}_{\mathrm{HQNS}} = 1160.8085$. However, the relative difference is small in practical terms: HQNS preserves $99.9908\%$ of the mean diversity score obtained by SA Parallel-11.

\begin{table}[t]
	\centering
	\caption{Solution quality comparison at $N=1000$ over ten independent runs.}
	\label{tab:quality_all_methods}
	\resizebox{\columnwidth}{!}{
		\begin{tabular}{lcccc}
			\toprule
			Method         & Mean $D$  & Std. Dev. $D$ & Mean Cost & Std. Dev. Cost \\
			\midrule
			Greedy         & 1160.7081 & 0.0000        & 64.2919   & 0.0000         \\
			SA Single      & 1160.8587 & 0.0424        & 64.1413   & 0.0424         \\
			HQNS           & 1160.8085 & 0.0336        & 64.1915   & 0.0336         \\
			SA Parallel-11 & 1160.9157 & 0.0143        & 64.0843   & 0.0143         \\
			\bottomrule
		\end{tabular}
	}
\end{table}

The results in Table 2 also clarify the relative positioning of the evaluated methods. SA Parallel-11 achieves the highest mean diversity score, followed by SA Single, HQNS, and Greedy. HQNS therefore should not be interpreted as the best method in terms of final objective value. Its relevance emerges from the quality--runtime trade-off: HQNS preserves nearly all of the diversity
score of the strongest parallel baseline while requiring substantially lower wall-clock time.

Relative to SA Parallel-11, the diversity retention ratio of HQNS is

\begin{equation}
	R_D =
	\frac{1160.8085}{1160.9157}
	\times 100\%
	=
	99.9908\%.
\end{equation}

The corresponding mean-cost increase is
\begin{equation}
	\Delta C =
	\frac{64.1915 - 64.0843}{64.0843}
	\times 100\%
	=
	0.1673\%.
\end{equation}

Thus, HQNS operates within a narrow solution-quality margin relative to the strongest classical baseline, while the resource analysis below shows that this near-baseline performance is obtained with substantially lower runtime and reduced computational footprint relative to SA Parallel-11.








\begin{figure}[t]
	\centering
	\includegraphics[width=\linewidth]{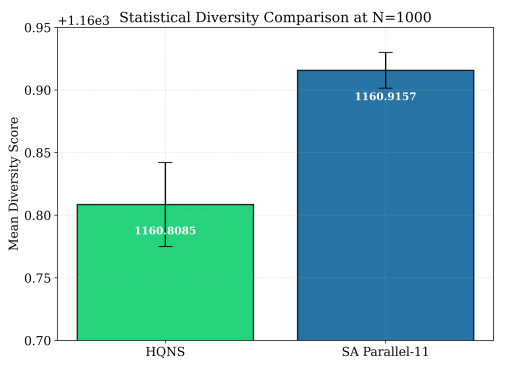}
	\caption{Statistical diversity comparison at $N=1000$ over ten independent runs. Error bars indicate one standard deviation. SA Parallel-11 achieves the higher mean diversity score, while HQNS preserves $99.9908\%$ of the SA mean diversity.}
	\label{fig:quality_retention_n1000}
\end{figure}

\subsection{Wall-Clock Time Reduction}

Although HQNS does not surpass SA Parallel-11 in mean diversity score, it substantially reduces total wall-clock execution time under the evaluated configuration. Table~\ref{tab:resource_all_methods} reports the mean resource usage over ten independent runs.

Greedy remains the fastest method, completing the $N=1000$ benchmark in $0.29 \pm 0.03$~s, but it also produces the lowest diversity score among the evaluated methods. SA Single achieves a slightly higher mean diversity score than HQNS with lower CPU and memory usage, but requires $37.49 \pm 0.70$~s. SA Parallel-11 achieves the best solution quality, but requires $75.32 \pm 3.06$~s and exhibits the highest CPU and memory footprint.

The most relevant comparison for HQNS is therefore against SA Parallel-11. This method represents the strongest classical baseline in terms of final solution quality. In this comparison, HQNS reduces wall-clock time from $75.32$~s to $3.84$~s, corresponding to

\begin{equation}
	\Delta_T =
	\left(
	1 -
	\frac{3.84}{75.32}
	\right)
	\times 100\%
	=
	94.91\%.
\end{equation}

This result is relevant from a deployment perspective because it shows that HQNS reaches near-SA-Parallel-11 solution quality with substantially lower total elapsed time. The reduction is not attributed only to QPU execution, but to the complete hybrid workflow, including frontier selection, reduced-QUBO construction, local variational optimization, and global solution update.






\begin{table}[t]
	\centering
	\caption{Computational resource comparison at $N=1000$ over ten independent runs.}
	\label{tab:resource_all_methods}
	\resizebox{\columnwidth}{!}{
		\begin{tabular}{lcccc}
			\toprule
			Method         & Time (s)         & Peak CPU (\%)      & Memory (MB)       & QPU Time (s)    \\
			\midrule
			Greedy         & $0.29 \pm 0.03$  & $15.75 \pm 8.60$   & $105.16 \pm 1.25$ & N/A             \\
			SA Single      & $37.49 \pm 0.70$ & $46.11 \pm 20.37$  & $205.94 \pm 2.50$ & N/A             \\
			HQNS           & $26.34 \pm 1.40$ & $483.00 \pm 22.96$ & $617.33 \pm 5.20$ & $6.76 \pm 0.32$ \\
			SA Parallel-11 & $75.32 \pm 3.06$ & $1097.03 \pm 4.81$ & $952.88 \pm 7.07$ & N/A             \\
			\bottomrule
		\end{tabular}
	}
\end{table}

\subsection{CPU and Memory Footprint}

The CPU and memory measurements further clarify the computational profile of the evaluated methods. Greedy and SA Single impose the lowest classical resource demands, but they do not represent the strongest quality-oriented baseline. SA Parallel-11, by contrast, achieves the highest mean diversity score by executing eleven independent annealing chains, which substantially increases CPU pressure and memory usage.

HQNS occupies an intermediate resource regime. Its peak CPU utilization is higher than Greedy and SA Single because the hybrid workflow includes frontier construction, variational optimization, GPU-assisted numerical operations, and QPU interaction. However, relative to SA Parallel-11, HQNS reduces peak CPU utilization from $1097.03\%$ to $387.00\%$, corresponding to a $64.68\%$ reduction. It also reduces peak memory usage from $952.88$~MB to $108.33$~MB, corresponding to an $88.61\%$ reduction.

These results should be interpreted carefully. HQNS is not the lowest-resource method among all baselines. Instead, it provides a lower-footprint alternative to the strongest quality-oriented classical baseline while preserving nearly all of its diversity performance. This distinction is essential for positioning HQNS as a resource-aware hybrid optimizer rather than as a universally dominant method.








\begin{figure}[t]
	\centering
	\includegraphics[width=\linewidth]{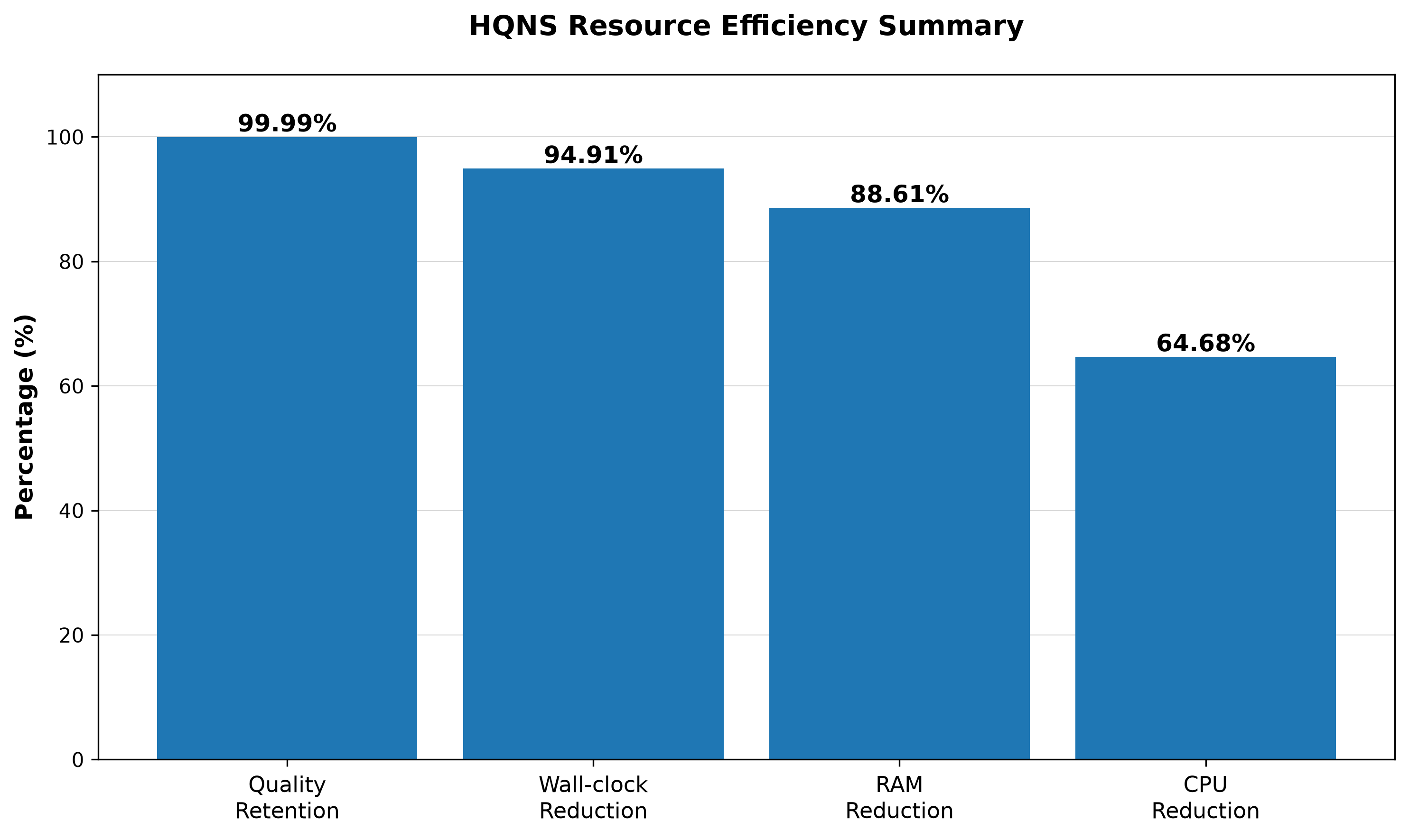}
	\caption{Quality-retention and resource-efficiency summary at $N=1000$. HQNS preserves $99.9908\%$ of the SA Parallel-11 mean diversity score while reducing wall-clock time, peak CPU utilization, and peak memory usage.}
	\label{fig:resource_efficiency_n1000}
\end{figure}


\subsection{QPU Execution-Time Stability}

A key feature of HQNS is that the quantum execution workload depends on the active frontier size $F$, rather than on the full problem dimension $N$. Table~\ref{tab:qpu_time_stability_access} reports QPU execution time across problem scales from $N=30$ to $N=1000$.

\begin{table}[t]
	\centering
	\caption{QPU execution-time stability across problem scales.}
	\label{tab:qpu_time_stability_access}
	\resizebox{\columnwidth}{!}{
		\begin{tabular}{lcc}
			\toprule
			Scale    & QPU Time (s) & Circuit Width (qubits) \\
			\midrule
			$N=30$   & 6.40         & 12                     \\
			$N=60$   & 7.53         & 12                     \\
			$N=120$  & 6.44         & 12                     \\
			$N=250$  & 6.45         & 15                     \\
			$N=500$  & 6.65         & 16                     \\
			$N=1000$ & 6.87         & 20                     \\
			\bottomrule
		\end{tabular}
	}
\end{table}

Despite a $33\times$ increase in global problem size from $N=30$ to $N=1000$, QPU execution time remains within an approximately $6$--$7.5$ second envelope. This behavior supports the central architectural premise of HQNS: the QPU workload is controlled by the reduced frontier size, not by the full QUBO dimension.

This does not imply that the complete hybrid workflow is independent of $N$. Classical preprocessing, frontier selection, and update operations still scale with the global problem size. However, the bounded QPU execution time demonstrates that HQNS enables the reuse of compact quantum circuits across increasingly large dense QUBO instances.

\begin{figure}[t]
	\centering
	\includegraphics[width=\linewidth]{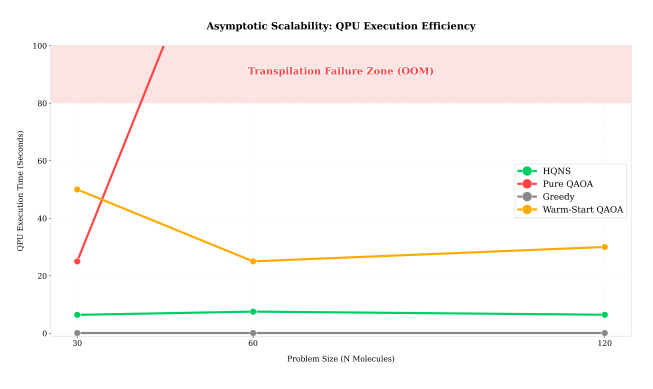}
	\caption{QPU execution-time stability across problem scales. HQNS maintains a bounded QPU execution envelope because each quantum stage operates on a compact active frontier rather than the full $N$-variable QUBO.}
	\label{fig:qpu_time_stability}
\end{figure}

\subsection{Quality--Resource Trade-off}

The combined results indicate that HQNS occupies a favorable quality--resource region. It does not maximize final diversity score relative to SA Parallel-11, but it preserves nearly all of the baseline quality while substantially reducing computational resource usage.

\begin{figure}[t]
	\centering
	\includegraphics[width=\linewidth]{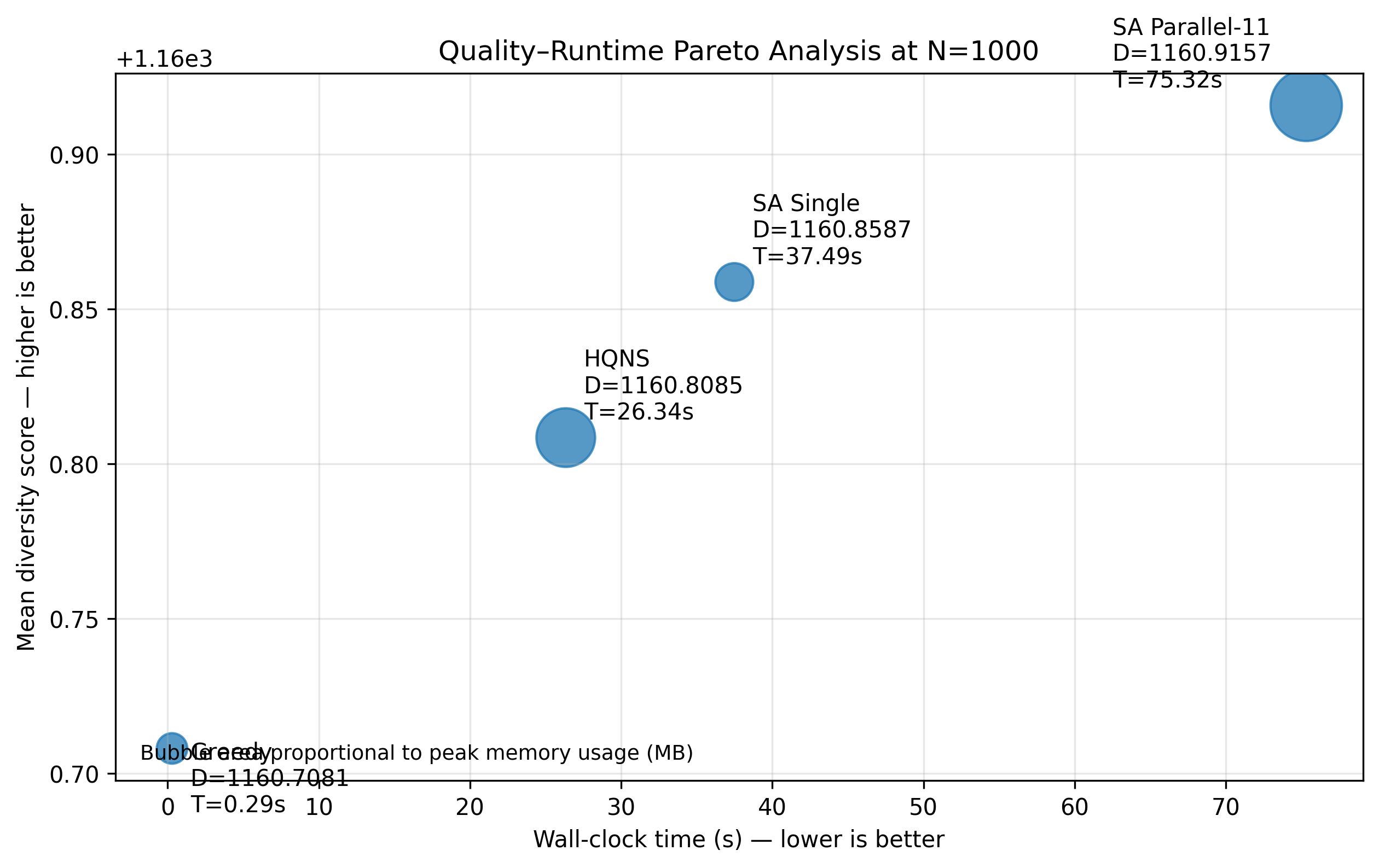}
	\caption{
		Pareto Analysis of Solution Quality versus Computational Cost at $N=1000$. The horizontal axis reports wall-clock execution time, while the vertical axis reports mean diversity score over ten independent runs. Bubble area is proportional to peak memory usage. HQNS occupies an intermediate operational region: it is substantially faster than SA Single and SA Parallel-11, while preserving near-baseline diversity relative to the strongest parallel classical baseline.
	}
	\label{fig:quality_runtime_pareto}
\end{figure}

To visualize the trade-off between solution quality and computational effort, Fig.~\ref{fig:quality_runtime_pareto} plots the evaluated methods in a quality--runtime space. The figure should not be interpreted as evidence that HQNS dominates all classical baselines. Greedy remains the fastest method and SA Parallel-11 remains the highest-quality method. Instead, the figure highlights the intermediate operational role of HQNS: it achieves near-parallel-SA diversity while substantially reducing wall-clock time.

This trade-off is important because practical optimization workflows are rarely selected using a single metric. Greedy is computationally inexpensive but has the lowest diversity score. SA Parallel-11 provides the strongest quality but requires the highest runtime and resource footprint. SA Single offers a strong low-resource classical compromise, whereas HQNS provides a hybrid alternative that prioritizes runtime reduction relative to SA baselines while maintaining a bounded quantum execution structure.

Table~\ref{tab:hqns_vs_sa_parallel_summary} summarizes the main trade-off indicators. These results suggest that HQNS is best understood as a resource-efficient hybrid optimization framework. Its value does not come from strict objective-value dominance over a strong classical baseline. Instead, it comes from preserving near-equivalent diversity performance while reducing the computational resource footprint and maintaining a bounded-width quantum execution structure.

\begin{table}[t]
	\centering
	\caption{HQNS trade-off relative to SA Parallel-11 at $N=1000$.}
	\label{tab:hqns_vs_sa_parallel_summary}
	\resizebox{\columnwidth}{!}{
		\begin{tabular}{lc}
			\toprule
			Metric                    & HQNS Relative to SA Parallel-11 \\
			\midrule
			Diversity score retained  & $99.9908\%$                     \\
			Mean cost increase        & $0.1673\%$                      \\
			Wall-clock time reduction & $94.91\%$                       \\
			Peak CPU reduction        & $64.68\%$                       \\
			Peak memory reduction     & $88.61\%$                       \\
			QPU execution time        & $6.76 \pm 0.32$ s               \\
			\bottomrule
		\end{tabular}
	}
\end{table}


This distinction is important for near-term quantum computing. Since present-day NISQ processors are not expected to provide unconditional advantage for dense large-scale QUBO optimization, practical utility may emerge from hybrid workflows that reduce resource pressure while integrating quantum subroutines into classical optimization pipelines.

\subsection{Summary of Findings}

The experimental results support four main observations. First, HQNS preserves $99.9908\%$ of the mean diversity score achieved by SA Parallel-11 at $N=1000$. Second, HQNS reduces wall-clock time by $94.91\%$, indicating a substantial acceleration of the full optimization workflow under the evaluated configuration. Third, HQNS reduces peak CPU utilization by $64.68\%$ and peak memory usage by $88.61\%$, demonstrating a lower computational resource footprint. Fourth, QPU execution time remains approximately stable across increasing problem scales, supporting the use of bounded-width quantum subproblems for large dense QUBO instances.

Together, these results indicate that HQNS provides resource-aware quantum utility rather than unconditional quantum advantage. It offers a practical pathway for incorporating near-term quantum processors into large-scale optimization workflows where solution quality, execution time, and computational resource consumption must be jointly considered.

\section{Discussion}

\subsection{Resource-Aware Quantum Utility}

The results presented in Section~VI indicate that HQNS should not be interpreted as a method that dominates strong classical heuristics in final objective value. In the $N=1000$ benchmark, SA Parallel-11 achieves a higher mean diversity score than HQNS. However, HQNS preserves $99.9908\%$ of the SA Parallel-11 mean diversity while reducing wall-clock time by $94.91\%$, peak CPU utilization by $64.68\%$, and peak memory usage by $88.61\%$.

The comparison across all four methods also clarifies the specific operating point of HQNS. Greedy is the lowest-resource method but produces the lowest diversity score. SA Single achieves a slightly higher mean diversity score than HQNS with lower CPU and memory usage, but requires longer wall-clock time. SA Parallel-11 achieves the highest mean diversity score but incurs the largest computational footprint. HQNS therefore should be interpreted as a hybrid runtime-efficient alternative to stronger SA baselines, rather than as a method that dominates every classical heuristic across every resource metric.

This distinction is central to the interpretation of the results. The practical value of HQNS lies not in unconditional quantum advantage, but in resource-aware quantum utility. In other words, HQNS provides a way to incorporate bounded-width quantum subroutines into large-scale optimization workflows while preserving near-baseline solution quality and reducing the computational footprint of the overall process.

This type of utility is particularly relevant in the NISQ era. Current quantum processors are not expected to solve dense large-scale QUBO problems through monolithic encodings, because such formulations require circuit widths, interaction graphs, and depths that exceed current hardware capabilities. HQNS addresses this limitation by restricting the quantum component to compact frontier subproblems, allowing the QPU workload to remain bounded even as the global problem size increases.

\subsection{Quality Versus Resource Consumption}

The quality--resource trade-off observed in this study is favorable for HQNS. At $N=1000$, HQNS does not achieve the best mean diversity score, but the loss in solution quality is extremely small relative to the reductions in computational resources. Specifically, HQNS preserves $99.9908\%$ of the SA Parallel-11 diversity score while reducing wall-clock time by $94.91\%$, peak CPU utilization by $64.68\%$, and peak memory usage by $88.61\%$.

This result suggests that strict objective-value comparison alone is insufficient for evaluating hybrid quantum--classical algorithms. In many practical environments, optimization methods are selected not only based on final objective value, but also based on runtime, hardware availability, memory pressure, scalability, and integration cost. A method that delivers nearly equivalent solution quality with substantially lower resource consumption may be operationally preferable, especially when optimization is embedded inside larger computational pipelines.

The small difference in mean diversity score should therefore be interpreted in context. The SA Parallel-11 baseline uses eleven independent annealing chains and obtains the best objective value in the evaluated configuration. HQNS, by contrast, reaches a near-equivalent solution-quality regime with a different computational structure: reduced frontier optimization, bounded quantum circuit width, and lower classical resource consumption.

\subsection{CPU and Memory Efficiency}

The reductions in peak CPU utilization and memory usage are among the most relevant results from a practical deployment perspective. SA Parallel-11 obtains strong solution quality by exploiting parallel stochastic exploration, but this increases CPU demand. HQNS reduces this pressure by limiting each local optimization step to a compact frontier and by avoiding repeated full-scale exploration of the dense QUBO during quantum-assisted refinement.

The observed memory reduction is also relevant. Dense QUBO and molecular similarity matrices can create substantial memory pressure as $N$ increases. In modern computing environments, memory availability is increasingly constrained by concurrent workloads such as machine learning, simulation, data processing, and large-scale analytics. A method that reduces memory consumption while preserving solution quality may therefore provide practical value even if it does not improve the best achievable objective value.

It is important to emphasize that this paper reports computational resource indicators, not direct electrical energy measurements. Lower CPU utilization, lower memory footprint, and shorter wall-clock time suggest the possibility of reduced energy demand under comparable hardware conditions, but direct energy claims require power-level measurements such as joules per run or watt-level telemetry. Therefore, energy efficiency is treated here as a potential implication of reduced resource usage, not as a directly measured result.

\subsection{Role of GPU Acceleration}

The HQNS workflow may use GPU acceleration for parallel numerical operations, such as statevector simulation or batched computations. However, the GPU is not the primary source of the observed resource-efficiency profile. In the evaluated configuration, HQNS relies on compact frontier subproblems, meaning that the size of the quantum-assisted optimization task remains limited even when the global molecular library grows.

This has two implications. First, HQNS does not require a high-end GPU to operate effectively under the evaluated setup. Second, the method remains conceptually compatible with heterogeneous computing environments in which CPU, GPU, and QPU resources are used selectively. The GPU can accelerate numerical operations, while the QPU handles bounded-width quantum subproblems and the CPU coordinates global solution updates.

This heterogeneous structure is consistent with the expected near-term deployment model for quantum computing. Rather than replacing classical computing, the QPU acts as a specialized accelerator inside a broader hybrid workflow.

\subsection{QPU Execution-Time Stability}

The stability of QPU execution time across problem scales is a central architectural result. Although the global problem size increases from $N=30$ to $N=1000$, the QPU execution time remains within an approximately $6$--$7.5$ second envelope. This behavior is expected because HQNS does not send the full $N$-variable QUBO to the quantum processor. Instead, each quantum execution operates on a frontier of size $F$.

This result should be understood in the context of what it would mean to attempt a monolithic QAOA encoding at the same scales. A direct QAOA formulation of the $N$-variable MDSSP QUBO would require a quantum register of $N$ qubits and $O(N^2)$ two-qubit gates per circuit layer to represent the dense pairwise interaction structure. For NISQ hardware such as superconducting transmon processors---which operate on planar or near-planar connectivity graphs---this immediately creates a transpilation bottleneck. Transpilation is the process of decomposing and routing abstract two-qubit gates onto the specific qubit-connectivity map of a physical device. In sparse hardware graphs, implementing a dense set of $O(N^2)$ long-range interactions requires SWAP chain routing that multiplies the number of physical CNOT gates by a factor proportional to the routing overhead. Beyond roughly $N>120$ for dense QUBO graphs, the resulting circuit depth routinely exceeds the coherence limits of current NISQ devices, and transpilation itself may fail to produce a valid executable circuit within practical compilation time or depth budgets. This region, in which both Pure QAOA and Warm-start QAOA effectively cease to be executable on real hardware, can be characterised as a \emph{Transpilation Failure Zone}.

HQNS is structurally immune to this barrier. Because each quantum stage is executed on an $F$-qubit subproblem with $O(F^2)$ interactions---where $F \in \{12, 15, 16, 20\}$ across all evaluated scales---the resulting circuits remain well within the connectivity, depth, and coherence limits of current NISQ processors regardless of the global problem size $N$. The observed QPU execution-time stability across the full $N=30$ to $N=1000$ range is therefore not incidental: it reflects the deliberate architectural choice to decouple the quantum circuit width from the global QUBO dimension through stochastic frontier decomposition.

This result supports the claim that HQNS decouples the quantum execution burden from the full global problem dimension. The complete hybrid workflow still includes classical operations that depend on $N$, such as frontier selection and solution update. However, the QPU component remains bounded by the frontier size. This is a key distinction between HQNS and monolithic QAOA, where the circuit width and dense interaction structure grow directly with the full problem size, eventually rendering transpilation and execution on current hardware infeasible.

From a practical perspective, this means that the same quantum execution architecture can be reused across increasingly large dense QUBO instances, provided that the frontier size remains within hardware-compatible limits. This makes HQNS suitable for near-term quantum workflows in which current QPUs are used as bounded local optimizers rather than full-scale global solvers.

\subsection{Why This Is Not Unconditional Quantum Advantage}

The results in this paper should not be interpreted as unconditional quantum advantage. HQNS does not demonstrate an asymptotic separation from classical algorithms, nor does it outperform the strongest classical baseline in mean solution quality. The SA Parallel-11 baseline remains slightly superior in final diversity score.

The contribution is different. HQNS demonstrates that a hybrid quantum--classical method can preserve near-baseline solution quality while using a resource-efficient computational structure that is compatible with current quantum hardware constraints. This is closer to practical quantum utility than to quantum advantage in the strict complexity-theoretic sense.

This distinction matters because overstating quantum advantage can weaken the credibility of near-term quantum optimization studies. The appropriate claim is that HQNS provides a resource-aware hybrid architecture for dense QUBO optimization, not that it proves superiority over all classical approaches.

\subsection{Implications for Practical Hybrid Optimization}

The findings suggest that hybrid quantum optimization should be evaluated using broader criteria than final objective value alone. For industrial and scientific applications, relevant metrics include execution time, memory footprint, CPU demand, hardware availability, stability across runs, and integration into existing computational pipelines.

HQNS is well aligned with this perspective. Its main advantage is the ability to restrict quantum execution to bounded-width subproblems while maintaining a competitive level of solution quality. This makes it potentially useful for settings where large dense QUBO instances must be solved repeatedly, such as molecular screening, portfolio selection, scheduling, and other subset-selection problems.

The framework also suggests a practical role for near-term QPUs. Instead of waiting for hardware capable of executing large monolithic quantum circuits, hybrid methods can use current devices as local optimizers embedded inside classical workflows. This may accelerate the development of practical quantum-assisted optimization before the arrival of fault-tolerant quantum computing.

\section{Conclusion}

This paper presented HQNS as a resource-efficient hybrid quantum--classical optimization framework for large-scale molecular diversity selection. Instead of treating near-term quantum processors as full-scale solvers for dense QUBO instances, HQNS uses the QPU as a bounded-width local optimizer embedded within a classical coordination workflow. By decomposing the global problem into stochastic frontier subproblems of size $F \ll N$, the method keeps the quantum execution workload compatible with NISQ hardware while preserving competitive solution quality.

The experimental results show that HQNS preserves $99.9908\%$ of the mean diversity score achieved by an 11-restart parallel Simulated Annealing baseline at $N=1000$. Although SA Parallel-11 obtains the higher mean final diversity score, HQNS substantially reduces computational resource usage, achieving a $94.91\%$ reduction in wall-clock time, a $64.68\%$ reduction in peak CPU utilization, and an $88.61\%$ reduction in peak memory usage under the evaluated configuration. These results indicate that HQNS is not a quality-dominant optimizer, nor is it universally resource-minimal across all classical baselines. Its contribution is more specific: HQNS reaches near-SA-Parallel-11 solution quality with substantially lower wall-clock time, CPU utilization, and memory footprint relative to the strongest quality-oriented classical baseline.

The QPU execution results further support the architectural motivation of HQNS. Across problem scales from $N=30$ to $N=1000$, QPU execution time remains within an approximately $6$--$7.5$ second envelope, suggesting that the quantum workload is primarily governed by the frontier size rather than by the full QUBO dimension. This bounded-execution behavior is essential for practical NISQ deployment, where circuit width, depth, connectivity, and noise accumulation remain major constraints.

The findings should not be interpreted as evidence of unconditional quantum advantage. HQNS does not demonstrate asymptotic superiority over classical algorithms, nor does it outperform the strongest classical baseline in mean solution quality. Its contribution lies instead in demonstrating resource-aware quantum utility: the ability to integrate near-term quantum subroutines into large-scale optimization workflows while preserving near-baseline solution quality and reducing computational resource demand.

Future work will pursue several concrete directions. First, a high-performance C++ implementation of the HQNS engine is planned to eliminate Python interpreter overhead and enable deployment in production-grade optimization pipelines. Second, experiments on alternative quantum backends---including IBM Heron and other superconducting architectures---will assess whether the quality--resource trade-off reported here generalizes across hardware platforms. Third, adaptive frontier-size strategies driven by local landscape curvature are under investigation, with the goal of dynamically balancing solution quality, circuit complexity, and resource usage during execution. Fourth, ML-assisted multi-crawling mechanisms are being explored, in which a machine-learning model guides frontier rotation rather than relying solely on stochastic marginal-impact scoring; this direction is expected to improve convergence speed and exploitation of the global solution landscape. Beyond the current study, future papers will investigate the application of HQNS to molecular subset selection at massively larger candidate libraries ($N > 1000$), targeting biodiversity score maximization in high-throughput virtual screening workflows. A dedicated evaluation under varying hardware configurations and larger fingerprint datasets will complement the resource-efficiency results presented here.

Overall, HQNS provides a practical pathway for deploying hybrid quantum optimization before the availability of fault-tolerant quantum computers. Its value lies not in replacing classical optimization, but in reorganizing the computational workload so that classical, GPU, and quantum resources can be combined more efficiently for large-scale combinatorial search.

More broadly, HQNS illustrates a practical path for
resource-aware quantum utility in the NISQ era.
Rather than pursuing monolithic quantum encodings
that exceed current hardware capabilities, bounded-width
hybrid decomposition strategies may provide a scalable
mechanism for integrating quantum processors into
real-world optimization workflows. This perspective
suggests that near-term value may emerge not from
unconditional quantum advantage, but from the efficient
orchestration of quantum and classical resources within
heterogeneous computing architectures.

\section*{Acknowledgment}

The authors gratefully acknowledge Gabriel Albuquerque, President and Founder of LACQ (Liga Acadêmica Nacional de Computação Quântica), for facilitating access to quantum computing resources and supporting the advancement of quantum research and education in Brazil.

The authors also thank IBM for providing access to quantum computing resources through the IBM Quantum Open Plan. The availability of these cloud-based quantum systems was essential for executing, validating, and benchmarking the hybrid quantum optimization strategies investigated in this work.

\bibliographystyle{IEEEtran}
\bibliography{references}

\begin{IEEEbiography}[{\includegraphics[width=1in,height=1.25in,clip,keepaspectratio]{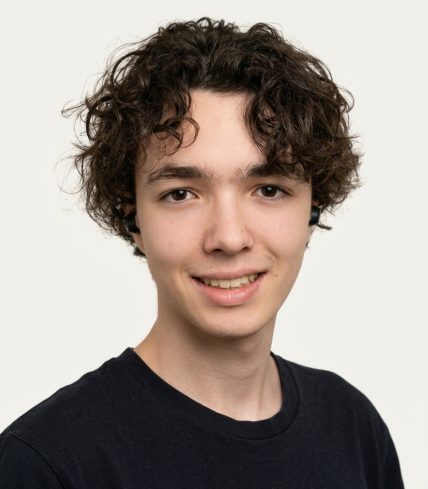}}]{{\textbf{NICOLAS MENDES DE ARAÚJO}}}

	is an independent researcher based in S\~{a}o Paulo, Brazil. His research interests focus on quantum computing, variational quantum algorithms, and stochastic optimization strategies for NISQ-era processors.

	He is the creator of the Hybrid Quantum Neighborhood Selection (HQNS) framework, specializing in mapping dense binary optimization problems and Quadratic Unconstrained Binary Optimization (QUBO) formulations onto constrained superconducting quantum architectures. His current work investigates frontier decomposition methods to bypass scalability barriers in heuristic quantum optimization, with applications in molecular diversity selection and complex combinatorial search spaces. ORCID: \url{https://orcid.org/0009-0006-0046-8801}

\end{IEEEbiography}

\begin{IEEEbiography}[{\includegraphics[width=1in,height=1.25in,clip,keepaspectratio]{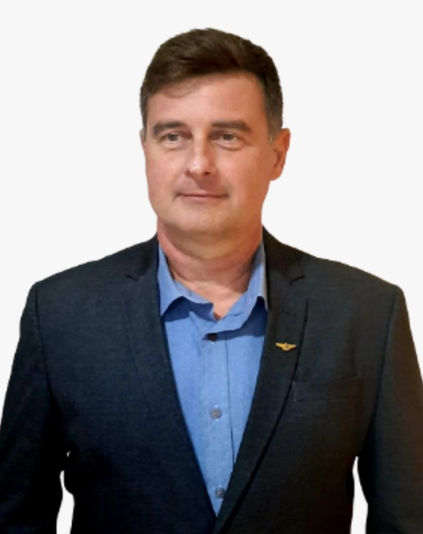}}]{{\textbf{LESTER DE ABREU FARIA}}}

	received the B.S., M.S., and Ph.D. degrees in Electronic Engineering from the Instituto Tecnol\'{o}gico de Aeron\'{a}utica (ITA), Brazil. He is a retired Colonel of the Brazilian Air Force and is currently affiliated with the Technological Institute of Aeronautics (ITA) as an associate professor. He is also conducting postdoctoral research in quantum technologies.

	His work spans electronics, sensing systems, artificial intelligence, and emerging quantum technologies. His current research interests include quantum algorithm engineering, hybrid quantum--classical computation, resource-aware evaluation of quantum algorithms, and advanced sensing technologies for aerospace and defense applications. He has participated in several national and international R\&D initiatives involving advanced technologies and aerospace systems.

\end{IEEEbiography}

\EOD

\end{document}